# Correlated fluctuations in spin orbit torque-coupled perpendicular nanomagnets

Punyashloka Debashis[1, 2], Rafatul Faria[2], Kerem Y. Camsari[2], Supriyo Datta[2], Zhihong Chen[1, 2]

[1]Birck Nanotechnology Center, [2]Department of Electrical and Computer Engineering

Purdue University, West Lafayette, USA

*Abstract* - Low barrier nanomagnets have attracted a lot of research interest for their use as sources of high quality true random number generation. More recently, low barrier nanomagnets with tunable output have been shown to be a natural hardware platform for unconventional computing paradigms such as probabilistic spin logic. Efficient generation and tunability of high quality random bits is critical for these novel applications. However, current spintronic random number generators are based on superparamagnetic tunnel junctions (SMTJs) with tunability obtained through spin transfer torque (STT), which unavoidably leads to challenges in designing concatenated networks using these two terminal devices. The more recent development of utilizing spin orbit torque (SOT) allows for a three terminal device design, but can only tune in-plane magnetization freely, which is not very energy efficient due to the needs of overcoming a large demagnetization field. In this work, we experimentally demonstrate for the first time, a stochastic device with perpendicular magnetic anisotropy (PMA) that is completely tunable by SOT without the aid of any external magnetic field. Our measurements lead us to hypothesize that a tilted anisotropy might be responsible for the observed tunability. We carry out stochastic Landau-Lifshitz-Gilbert (sLLG) simulations to confirm our experimental observation. Finally, we build an electrically coupled network of two such stochastic nanomagnet based devices and demonstrate that finite correlation or anti-correlation can be established between their output fluctuations by a weak interconnection, despite having a large difference in their natural fluctuation time scale. Simulations based on a newly developed dynamical model for autonomous circuits composed of low barrier nanomagnets show close agreement with the experimental results.

**Introduction**

When the energy barrier separating the two stable states ("UP" and "DOWN") of a nanomagnet is comparable or smaller than the ambient thermal noise, its magnetization fluctuates randomly between the two states. One of the simplest applications that harnesses this inherent stochasticity of a low barrier nanomagnet (LBNM) is true random number generators (TRNG)[1–3]. However, the full potential of LBNM based hardware can only be realized when the probability of the LBNM magnetization being in the "UP" or "DN" state is tunable by an external input. Such a hardware has been given the term 'p-bit', which stands for probabilistic bit[4,5]. Being essentially tunable random number generators, p-bits have recently been shown as natural hardware accelerators for unconventional computing tasks such as Ising computing[6,7], Bayesian networks[8,9], neural networks[10,11] and invertible Boolean logic[4].

Several implementations of LBNM based TRNGs have been demonstrated in the last few years, while only a few included the output tunability. One such device is based on an SMTJ with an in-plane LBNM as the free layer, where the tunability of the output state is obtained through STT[1]. It is well known that the major reliability issue in STT-MRAM is the result of the read and write operations sharing the same access path through the entire MTJ stack. Furthermore, having a common read and write path does not allow for the isolation of the input and output signals, and hence makes it difficult to concatenate these devices into a network. Therefore, a three-terminal device with SOT based output tunability is much more desirable due to the separation of the write current path from the read current path[12]. Such devices have been proposed for in-plane LBNMs[13–15]. However, recent simulation studies suggest that a dense array of in-plane LBNMs have significant magnetic dipolar interactions[1]. Such interactions could lead to compromised randomness and unwanted correlations between SMTJs in a large network. Moreover, SOT tunability of in-plane magnetization occurs through the so-called anti-damping mechanism, which is energy inefficient since it must overcome a large demagnetization field[16]. Therefore, LBNMs with perpendicular magnetic anisotropy (PMA) are ideal for high density, smaller pitch size arrays that are essential for large network implementations. However, current material systems that exhibit SOT can only generate spins with in-plane polarization, which is orthogonal to the magnetization of the low barrier PMA magnet, hence, complete tunability of its output state is not possible through SOT alone[17].

In this work, we demonstrate for the first time, an SOT tunable random number generator made of a PMA LBNM. The SOT tunability is realized through a small tilt in the magnetic anisotropy axis, as is evidenced by our experiments and supported by sLLG simulations. We then couple two such devices via electrical connections and study the correlation in their output fluctuations. Our experiments show that a weak coupling strength, that is 10 times smaller than the critical current required for deterministic switching, is sufficient to establish correlations between the outputs of the two devices. By changing the connection polarity, we show that the correlation can be changed from positive to negative. Our studies also show that two LBNMs with different time scales of fluctuation can get correlated efficiently. We perform simulations on this coupled 2 p-bit system using a dynamical model of autonomous circuits with all the required parameters taken from experiments. The simulation results show good matching with the experiments. This demonstration of a novel tunable TRNG and its behavior in an electrically coupled network provides important insights towards realizing large p-bit networks for unconventional computing tasks.

**Designing low barrier nanomagnets with perpendicular magnetic anisotropy**

The thermal stability factor of a nanomagnet is given by $E_B/k_BT$, where $E_B = K_{eff}V/2$ is the energetic barrier separating the two stable magnetization states. Here, $K_{eff}$ is the effective anisotropy energy density and V is the volume of the nanomagnet. When $E_B$ is comparable to the ambient thermal energy $k_BT$, the magnetization randomly fluctuates between the two stable states, thus realizing a "stochastic nanomagnet".

We first engineer $K_{eff}$ of our magnetic material (CoFeB) by varying the thickness ($t_{CoFeB}$) of the deposited PMA films, shown in Fig. 1 (a). The anisotropy of such a stack is given by[18]:

$$K_{eff} = K_i/t_{CoFeB} - M_S^2/2\mu_0 \quad \text{...............................................................................} (1)$$

arising from the competition between the interface anisotropy ($K_i$) and the demagnetization ($M_S^2/2\mu_0$) [ref]. We then follow the method used by Hayashi et al.[19] to characterize $K_{eff}$ of our stacks, as shown in Fig. 1 (b). For films with out-of-plane anisotropy, the obtained anomalous Hall resistance ($R_{AHE}$) in the presence of an in-plane magnetic field is fitted with a second order curve to obtain $H_K$ (where $H_K$ is the effective anisotropy field, given by $K_{eff}/M_S$). For films that have net in-plane anisotropy, $H_K$ is obtained through a linear fit of the $R_{AHE}$ vs. out-of-plane field curve. We observe a clear decreasing trend of $K_{eff} \times t_{CoFeB}$ vs. $t_{CoFeB}$ for as deposited films as well as samples annealed at 250 °C for one hour, as shown in Fig. 1 (c). The annealed stack with $t_{CoFeB}$=1.3 nm, corresponding to the lowest $K_{eff}$, is then chosen to fabricate the stochastic nanomagnet devices. The fabricated devices consist of lithographically defined PMA nanomagnets with a diameter of 100 nm on top of tantalum (Ta) Hall bars, as shown in Fig. 2 (a). The combination of low $K_{eff}$ through the thickness optimization and low volume through the lithography patterned small diameter results in a small $E_B$ at room temperature. Consequently, these uniquely designed stochastic nanomagnets fluctuate randomly between the "UP" and "DN" magnetic states as depicted in the cartoon in the top right inset of Fig. 2 (a). This random fluctuation is electrically read out through the anomalous Hall effect (AHE), giving the random telegraphic signal as the output, shown in Fig. 2 (b). The magnetization dwell time in the "UP" and the "DN" state forms a distribution that is well fitted by an exponential envelope, which suggests that the fluctuation is a random Poisson process[1]. To further test the quality of the randomness, we perform evaluations using the standard statistical test suite provided by the National Institute of Standards and Technology (NIST)[20]. The generated bit stream by our device passed all 9 tests that were performed, showing cryptographic quality randomness (test results in the supplementary information).

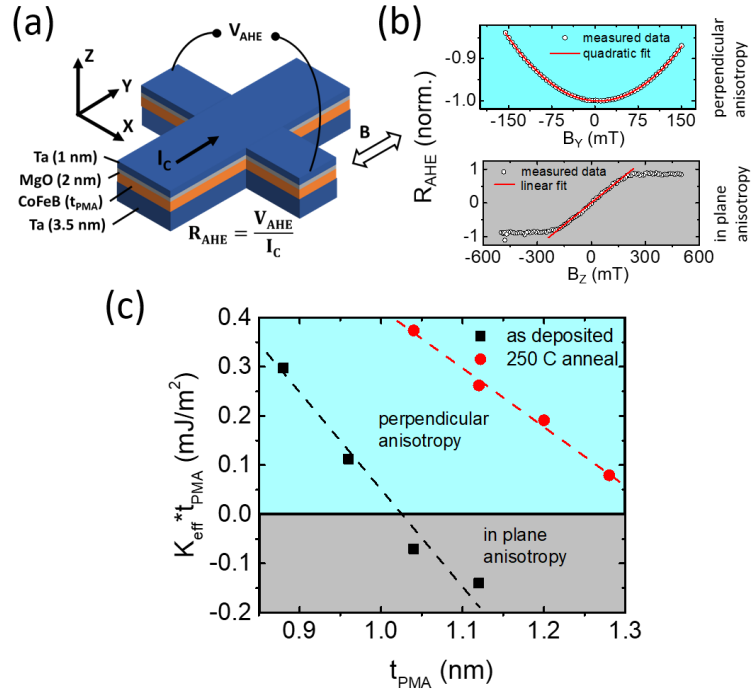

Figure 1: (a) PMA stack with varying CoFeB thickness. (b) Anomalous Hall resistance as a function of magnetic field along the hard axis. For a PMA magnetic stack, the field is applied in the in-plane direction and the measured data are fitted with a parabolic curve to extract the effective anisotropy field ($H_K$). For an IMA magnetic stack, the field in applied perpendicular to the plane and the resultant plot is fitted with a linear fit to extract $H_K$ (c) $K_{eff} \times t_{PMA}$ as a function of CoFeB layer thickness before and after 250 °C anneal.

## Tunability of the random output through spin orbit torque

We demonstrate that the mean value of the random numbers can also be tuned by a DC current through the giant spin Hall effect (GSHE) Ta Hall bar, as shown in Fig. 2. Fig. 2 (a) shows the measurement configuration, where a constant DC current ($I_{DC}$) through the GSHE underlayer is applied on top of a small AC read current ($I_C$). As shown in Fig. 2 (d), depending on the sign and magnitude of $I_{DC}$, the magnetization direction favors the "UP" or "DN" direction, resulting in the sigmoidal curve for the average. We call this device a probabilistic bit, i.e. a p-bit. Representative signals at three different $I_{DC}$ are presented in the panels to the right of Fig. 2 (d). The effect of DC current can also be seen by plotting the dwell time in "UP" and "DN" states for various $I_{DC}$, as shown in Fig. 2 (e) (measurement done at 250 K). This modification in the dwell time directly results in the tunability of the average magnetization. It is worth mentioning here that the small read current can also affect the state of the output, especially for a LBNM having a thermal barrier close to zero, and hence has to be carefully mitigated by design. This read disturb issue however is negligible in our case, where the energy barrier of the LBNM is ~18 $k_BT$.

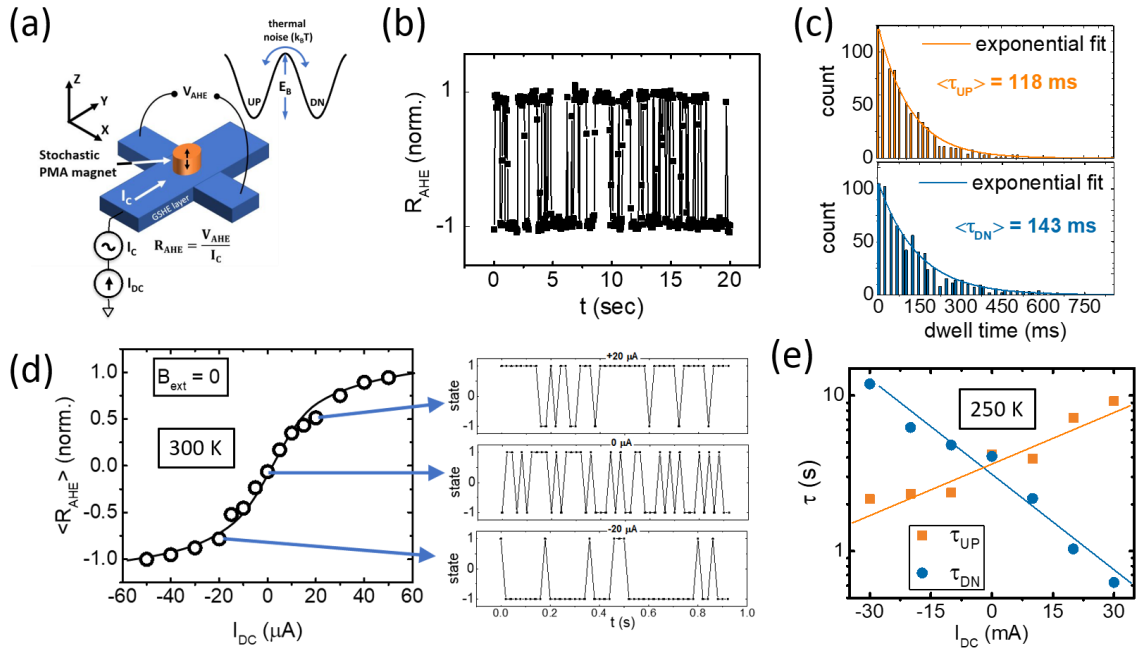

Figure 2: (a) Schematic of the measurement configuration with the fabricated device using the optimized $t_{PMA}$ giving the lowest $H_K$. The magnetic island has a diameter of ~100 nm. Cartoon representing the energy diagram of the perpendicular magnetization is shown in the top right inset. The two states, i.e, "UP" and "DN" are separated by a small energy barrier $E_B$, so that thermal energy is sufficient to randomly fluctuate the magnetization between the two states. (b) Measured anomalous hall resistance for a fixed small read current ($I_C$) and no DC current ($I_{DC}$). The random telegraphic signals arise from the random fluctuations of the perpendicular magnetization between "UP" and "DN" states. (c) Histogram of the dwell time in "UP" and "DN" states. Both histograms are well fitted by an exponential envelope, showing that the magnetization flipping can be represented by a random Poisson process. The average dwell time ($\tau_{UP}$ and $\tau_{DN}$) are calculated from the exponential fit. (d) Measurement with a DC charge current through the GSHE underlayer to obtain tunability. A sigmoidal curve is obtained for the average $R_{AHE}$ vs. $I_{DC}$, showing tunability for a PMA LBNM without any external magnetic field. Each point on this curve is obtained by averaging the random telegraphic output, representative data sets shown in the three panels on the right. (e) The dwell times in "UP" and "DN" state changes as a function of $I_{DC}$, which leads to the sigmoidal curve for average magnetization state.

*Physics of tuning low barrier PMA magnet through in-plane spins*

Since the polarization direction of the generated spins due to the charge current through the GSHE underlayer lies in the X-Y plane, it is surprising to see a tunability of the perpendicular magnetization by the DC current. An obvious hypothesis that we first considered was: the Oersted field generated by $I_{DC}$ points along the Z-axis at the edges of the hall bar, and could potentially favor one magnetization state over the other, leading to the observed tunability. This is illustrated in Fig. 3 (a). To test this hypothesis, we measure the magnetization response as a function of an applied magnetic field along the Z-direction. As expected, the average magnetization shows a sigmoidal behavior, as this external field favors the "UP" direction for positive field values and "DN" direction for negative field values. We then repeat the same measurement in the presence of a constant $I_{DC}$. Any constant Oersted field in the Z direction produced by this $I_{DC}$ would result in a horizontal shift or offset of the sigmoidal response, by an amount equal to the average magnetic field exerted on the magnet along Z-direction due to $I_{DC}$. We measure this shift, "$B_{offset}$" for various $I_{DC}$ and plot $B_{offset}$ vs. $I_{DC}$ in the right graph of Fig. 3 (b). There are two observations from this graph that contradict the hypothesis of the Oersted field induced tunability. Firstly, $B_{offset}$ is not a linear function of $I_{DC}$, which is different from the Oersted field linearly following the current, $B = \mu_0 I_{DC}/2W$. It can be clearly seen that $B_{offset}$ saturates for $|I_{DC}| > 10$ μA. Secondly, in the region where $B_{offset}$ is linear with $I_{DC}$, the slope, $B_{offset}/I_{DC} = 4 \times 10^{-1}$ mT/μA, is much larger than the expected value of $\mu_0/2W = 3 \times 10^{-3}$ mT/μA for the case of current induced Oersted field.

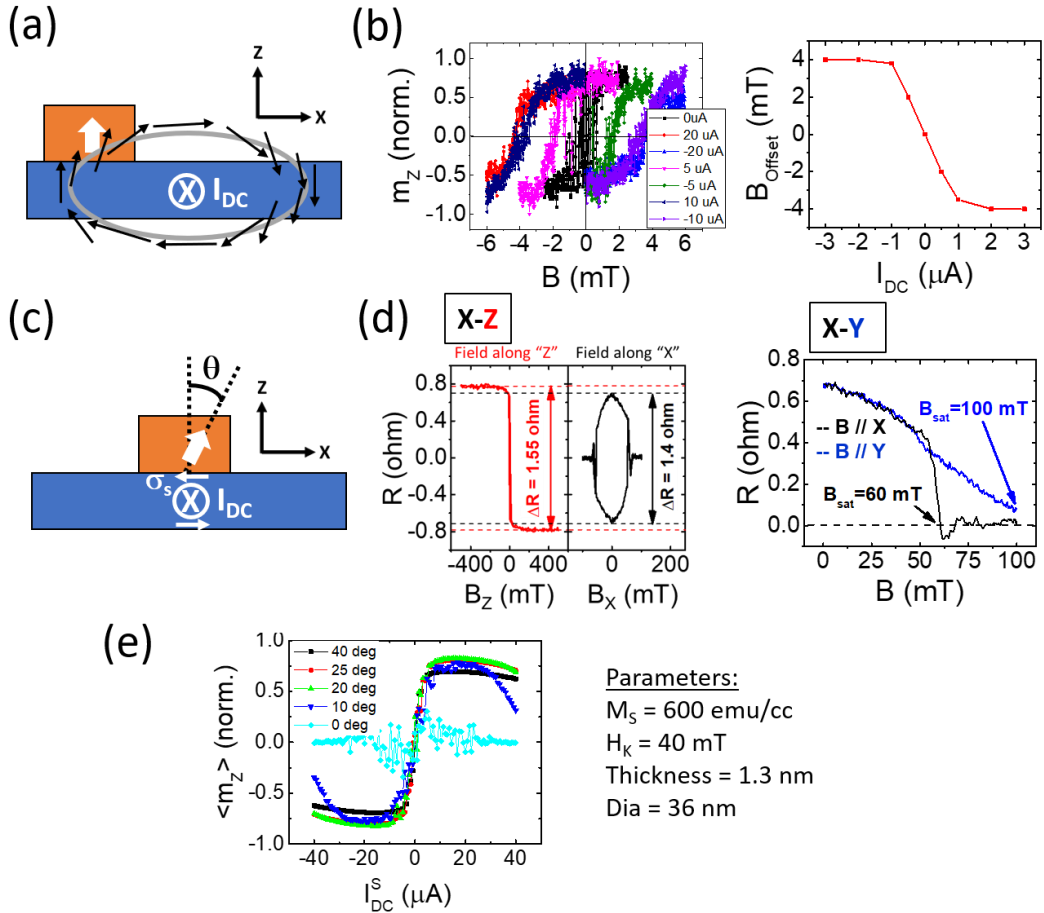

Figure 3: (a) Possible explanation of the obtained tunability. A lithographic misalignment could lead to the magnet island being situated towards one edge of the Ta electrode, where the perpendicular component of the Oersted field due to the charge current could lead to the observed tunability. (b) Device output as a function of external magnetic field in the Z-direction, in the presence of DC current through the GSHE underlayer. Offset field ($B_{offset}$) due to $I_{DC}$ is obtained from the horizontal shift in the output curves. The plot on the right shows $B_{offset}$ vs. $I_{DC}$, which clearly displays a saturating behavior. Also, the slope in the linear region is more than an order of magnitude larger than that expected from the Oersted field. (c) Another possible explanation of the obtained tunability. A tilted anisotropy in the nanomagnet leads to a non-zero $m_x$ component of the magnetization that can be tuned by the spin current through the GHSE underlayer. Due to the tilted anisotropy field, tuning $m_x$ by the in plane spin currents leads to tuning $m_z$. (d) Measured anomalous Hall signal as a function of magnetic field in X, Y and Z direction. From the X-Z plot, we can deduce the tilt angle $\theta$ from the ratio of saturation signal. From the X-Y plot, we notice that it is easier to saturate the magnetization in plane in the X direction compared to the Y direction, suggesting that the tilt of magnetization lies in the X-Z plane. (e) sLLG simulations of the above device with an applied DC charge current for various magnetization tilt angles. The charge current flows in the Y direction in the GSHE underlayer, producing spins with polarization along X direction that are responsible for the observed tunability.

We therefore hypothesize that a slightly tilted magnetic anisotropy direction is responsible for the observed tunability, as illustrated in Fig. 3 (c). Essentially, if the magnetization tilt is in the X-Z plane, it can lead to X-polarized spins in the SOT underlayer favoring one state over other. A +X directed spin will favor the (X, Z) quadrant for the magnetization, resulting in a positive $m_Z$, which will register as an "UP" in the AHE measurement (since $R_{AHE}$ is proportional to $m_Z$). On the other hand, a –X directed spin will favor the (-X,-Z) quadrant, resulting in a negative $m_Z$ or "DN" direction for the magnetization. This is similar to the engineered tilted anisotropy work by You et al.[21] applied on stable, large barrier magnets. In our case, since the magnetic stack is designed to have a very low perpendicular anisotropy, any small in-plane anisotropy developed during the film deposition process can lead to a significant tilt angle that is otherwise undetectable in magnets with strong perpendicular anisotropy. To test out this hypothesis, we carry out AHE measurements as a function of external magnetic fields along Z, Y and X directions on another device made of the same stack. Firstly, we carry out $R_{AHE}$ vs. $B_Z$ measurements as shown in Fig. 3 (d) left plot. It can be seen that the saturation value of $R_{AHE}$ is noticeably larger than the remanent value. The tilt angle, θ can be estimated by the relation θ = $\cos^{-1}(R_{AHE, remanent}/R_{AHE, saturation})$ as is done by You et al.[21]. From the measured data, θ is estimated to be around 25 degrees. Next, we carryout $R_{AHE}$ measurements in the present of $B_X$ (black curve) and $B_Y$ (blue curve), shown in Fig. 3 (d) right plot. In these measurements, the $R_{AHE}$ saturates to a zero value for large applied fields (B > $B_{sat}$) since the perpendicular component of the magnetization vanishes as the magnetization is progressively forced to lie in the X-Y plane. By comparing $B_{sat}$ for the X directed field and Y directed field, it is seen that the magnetization can be forced along the X direction more easily than the Y direction, as $B_{sat, X} < B_{sat, Y}$. This suggests that the tilt direction is in the Z-X plane, toward the X-axis.

*Stochastic LLG simulations of the magnetization dynamics with tilted anisotropy*

The feasibility of SOT based output tunability of a p-bit made of a low barrier PMA magnet with tilted anisotropy is confirmed by numerically solving the stochastic Landau-Lifshitz-Gilbert (sLLG) equation with a monodomain macro-spin assumption:

$$(1 + \alpha^2)\frac{d\hat{m}}{dt} = -|\gamma|\hat{m} \times \vec{H} - \alpha|\gamma|\hat{m} \times \hat{m} \times \vec{H} - \frac{1}{qN_s}\hat{m} \times \hat{m} \times \vec{I_s} + \frac{\alpha}{qN_s}\hat{m} \times \vec{I_s} \quad \ldots(2)$$

where, $\vec{H}$ is the total effective field including the anisotropy field $\vec{H_k}$ along a direction $\theta^0$ tilted with respect to the $Z$ axis on the $X - Z$ plane and the three dimensional uncorrelated thermal noise field $\vec{H_n}$ having Gaussian distribution with mean of $\langle H_n \rangle = 0$ and standard deviation of $\langle H_n^2 \rangle = 2\alpha kT/|\gamma|M_s V$, $\vec{I_s}$ is the spin current polarized along the $X$ direction, $N_s = M_s V$ is the total magnetic moment with $M_s$ being the saturation magnetization and $V$ being the volume of the magnet, $\alpha$ is the damping coefficient, $\gamma$ is the gyromagnetic ratio. Magnet parameters used in the simulation are: $H_k = 400$ Oe, $M_s = 600$ emu/cc, diameter $D = 36\ nm$, thickness $t = 1.3\ nm$, $\alpha = 0.1$. The magnetization fluctuates in time owing to its low thermal barrier. The average magnetization component in the Z direction (<$m_Z$>) is plotted as a function of DC spin current in Fig. 3 (e). When the magnet's anisotropy does not have any tilt with respect to the Z direction, then the in-plane spins do not affect preferred direction of $m_Z$. Hence, the average of $m_Z$ stays around zero. However, for tilt angles larger than 10 degrees, complete tunability of $m_Z$ can be obtained, as can be seen in Fig. 3 (e).

Note that at high currents the sLLG simulation suggests that the z component of the magnetization can completely vanish as the magnetization is pulled into the ±X direction. We believe that these currents are very large and experimentally not accessible in our system.

**Electrically coupled network of two p-bits**

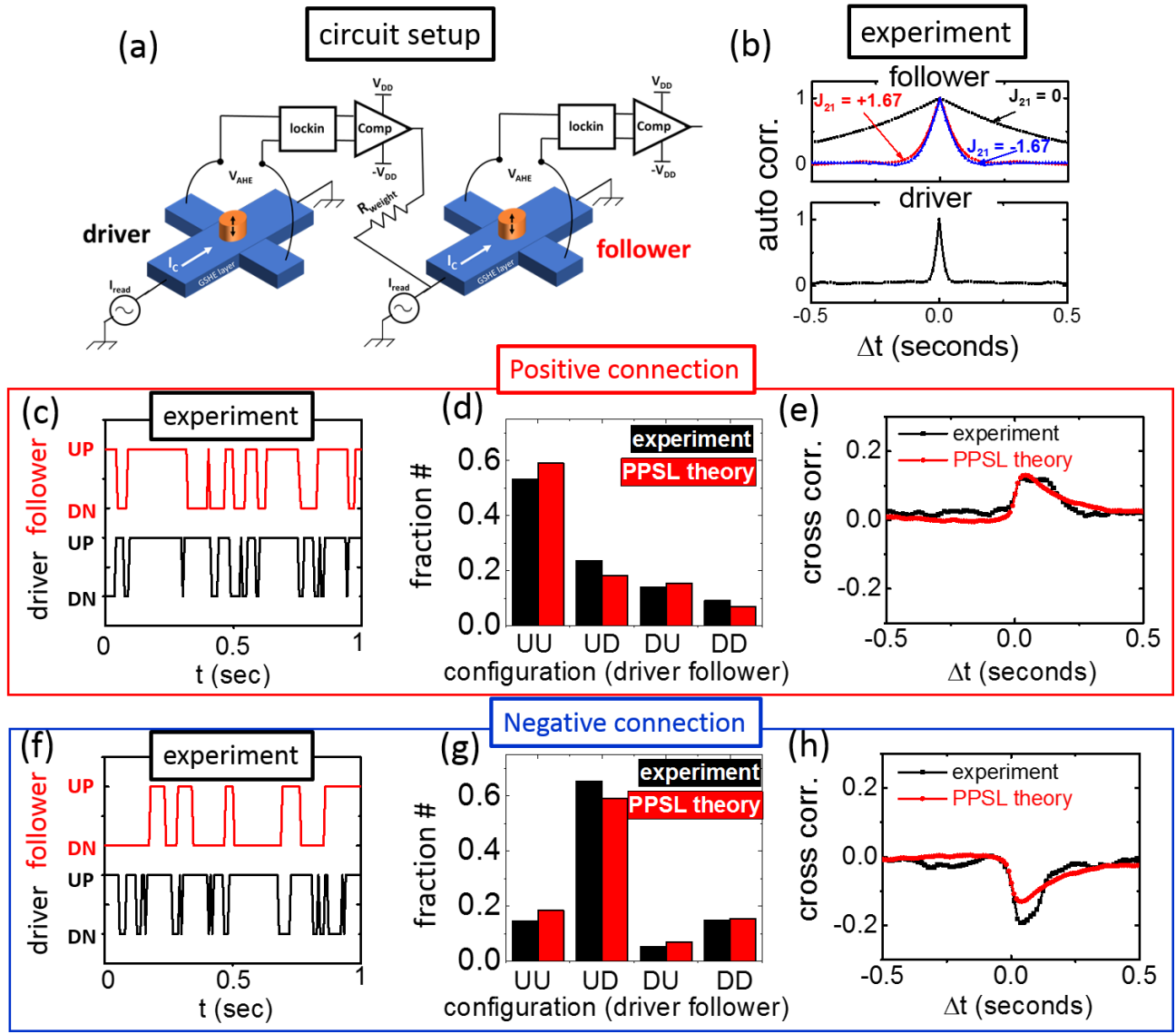

Figure 4: (a) The circuit to implement directed connection between two p-bits. (b) Auto correlation of the outputs of the driver (bottom) and the follower (top) for different connection configurations. Follower p-bit is much slower than driver for no connection case, but starts to respond faster when positive or negative connection is established between the two p-bits. (c) Time traces of the two p-bits. With positive connection established between them, positive correlation starts appearing, which is also seen from by plotting the histograms of the four possible states in (d). The parallel configurations (UU) are more frequent. This is closely matched by PPSL simulations. (e) The "relatedness" between the driver and follower signals is quantified by the cross correlation, which shows a positive peak. The correlation coefficient given by the height of the peak and the time scale of the correlation, given by the FWHM of the peak are both closely matched by PPSL simulations. (f) (g) and (h) are for the case of negative correlation.

*Experimental results*

In this section, we study two electrically coupled p-bits. The stochastic output of the first p-bit ("driver") is amplified and provided as the input to the GSHE underlayer of the next p-bit ("follower"). The amplification is done by SRS 830 lockin amplifiers, with an averaging time of 3 ms, which is much smaller than any other time scale in the experiment. The outputs of the lockin amplifiers are then fed to comparators in order to digitize the signals. We vary the connection configuration between the two devices to observe their behaviors. Fig. 4 (a) illustrates the circuit setup. We study three configurations: no connection ($R_{weight}$ = infinity), positive connection ($R_{weight}$ = 400 KΩ and amplifier gain is positive) and negative connection ($R_{weight}$ = 400 KΩ and amplifier gain is negative). Fig. 4 (b) shows the autocorrelation of the output signals of the two devices ($A_{driver}$ and $A_{follower}$) for the three configurations, obtained by the following formula:

$$A_{driver}(\Delta t) = \sum_{t=0}^{T-\Delta t}(X_t - \bar{X})(X_{t+\Delta t} - \bar{X}) ; \; A_{follower}(\Delta t) = \sum_{t=0}^{T-\Delta t}(Y_t - \bar{Y})(Y_{t+\Delta t} - \bar{Y})$$

In the above formulae, $X_t$ and $Y_t$ are the state of the driver and follower respectively at time $t$; $\bar{X}$ and $\bar{Y}$ are the respective mean values; $T$ is the total measurement time. When unconnected, the two devices fluctuate at different time scales, evidenced by the markedly different full width at half maximum (FWHM) of their autocorrelation peaks. However, when either a positive or negative connection is established between the two devices, the fluctuation time scale of the follower becomes closer to that of the driver, seen again by the FWHMs.

Fig. 4 (c) and (f) show representative sections of the time traces of the output signals of the driver and the follower for positive and negative connection respectively, where the emergence of positive and negative relation can be observed. To quantify the relatedness between the two signals, the histogram of the four possible configurations are plotted in Fig. 4 (d) and (g). It is observed that the driver signal has some inherent bias towards the UP state, possibly due to the presence of an unwanted magnetic field in the measurement chamber. Therefore, to accurately quantify the relatedness between the two outputs, we calculate the cross correlation between the two signals ($C_{driver,follower}$), obtained by introducing a relative time shift ($\Delta t$) between the two output signals and calculating the inner product as a function of this shift according to the following formula:

$$C_{driver,follower}(\Delta t) = \sum_{t=0}^{T-\Delta t}(Y_t - \bar{Y})(X_{t+\Delta t} - \bar{X})$$

This metric is less prone to the inherent bias as the correlations are calculated from signals after subtracting their respective mean values. Also, this metric preserves the time dependence of the relatedness. Any misleading relatedness observed in the histogram due to inherent biases in the two signals would not have time dependence, and hence would not contribute to the peak structure on the cross correlation plots.

We plot the normalized cross correlation and autocorrelations obtained from the following normalization:

$$\hat{C}_{driver,follower}(\Delta t) = \frac{C_{driver,follower}(\Delta t)}{\left(A_{driver}(0) \times A_{follower}(0)\right)^{1/2}}$$

$$\hat{A}_{driver}(\Delta t) = \frac{A_{driver}(\Delta t)}{A_{driver}(0)} ; \; \hat{A}_{follower}(\Delta t) = \frac{A_{follower}(\Delta t)}{A_{follower}(0)}$$

The black curves in Fig. 4 (e) and (h) show $\hat{C}_{driver,follower}$. The correlation coefficient is just the peak value of $\hat{C}_{driver,follower}$. From the above analysis we obtain the following insights for the different connection configurations.

*No Connection:* For the no connection case ($R_{weight}$ = infinity), the outputs of the two devices are essentially two independent random bit streams. An important finding from this experiment is that the two unconnected p-bits here have markedly different time scales of fluctuation, as is seen by full width at half maximum (FWHM) of the auto correlation plots for driver and follower signals. The driver fluctuates at a faster time scale, with an FWHM = 24 ms, whereas that for the follower p-bit is much slower with a FWHM of 648 ms.

*Positive/Negative connection:* Next, we use $R_{weight}$ = 400 KΩ and choose the connection polarity to implement a positive correlation. The choice of $R_{weight}$ and $V_{DD}$ together result in a current of 25 μA input to the second device, which is smaller than the critical current required for deterministic switching of the magnetization direction. Considering a magnet with an energy barrier of $E_B \approx 15$ $k_B T$, spin Hall angle of Tantalum, $\theta_{Ta} = 0.07$ and the Hall bar width of W =200 nm, the critical switching current without thermal assistance can be calculated to be $I_{critical}$ ~300 μA from the formula given by Liu et al.[22]. Therefore, the current required here for establishing a correlation between the two p-bits is more than 10 times smaller than $I_{critical}$. The effect of a positive connection can be seen in the time traces, where the follower's output signal weakly follows that of the driver, while showing intermittent random flips. The cross-correlation peaks around Δt = 0 and dies off with a FWHM =162 ms, suggesting that the follower p-bit responds to the input provided by the driver in the time scale of the driver. It is also interesting to see that the follower, which was much slower than the driver in the unconnected case, starts to respond with a speed close to that of the driver for the positive connection case. This is quantified by the follower FWHM decreasing to 100 ms, as shown in the figure inset. Similarly, for the negative connection case, a negative peak in the cross correlation can be seen around Δt = 0. The speed of the follower becomes closer to that of the driver, as quantified by the reduction in FWHM of the follower to 75 ms.

*Autonomous PSL simulations*

Unlike inherently synchronous digital platforms, the hardware proposed in this article is completely autonomous without any sequencers to enforce any specific update order. To model this autonomous hardware, we have used a simple behavioral model as described by Sutton et. al[23] that is benchmarked against coupled stochastic Landau-Lifshitz-Gilbert (sLLG) equation for capturing low barrier nanomagnet physics. In this model, each p-bit in the network flips with a probability of $p$ controlled by the input $I_i$ described by the following equation:

$$m_i(t + \Delta t) = m_i(t) \times sgn(exp(-p_i) - r_{[0,1]})$$

$$p_i = \frac{\Delta t}{\tau_{corr,i}} exp(-m_i(t) I_i(t))$$

where, $m_i$ is the output state of the i-th p-bit, $\Delta t$ is the simulation time step, $r_{[0,1]}$ is a random number between 0 and 1, $\tau_{corr,i}$ is the correlation time of the p-bit under zero input.

The interconnection of the p-bits are described by the following synapse equation:

$$I_i(t) = \sum_j J_{ij} m_j(t) + h_i$$

Where, $J_{ij}$ is the dimensionless coupling term obtained from the experimental parameters using the following mapping:

$$J_{ij} = \frac{V_{DD}}{R_{weight,ij} I_{DC,0}}$$

Where, $I_{DC,0}$ is the tanh fitting parameter for the sigmoidal response of the follower.

Experimentally obtained parameters used in the PPSL simulation are: $I_{DC,0} = 15\ \mu A$, $V_{DD} = 10\ V$, $R_{weight,ij} = 400\ K\Omega$, $\tau_{corr,1} = 24\ ms$ $\tau_{corr,2} = 648\ ms$.

$h_j$ is the bias term provided to the j-th p-bit. The fractional occupation of the driver p-bit in the "UP" state that is obtained from the experimental histograms gives $h=0.63$ for the driver. For the follower, obtaining $h$ is not straight forward. However, for our experiment, $h=0$ fits the measurement data nicely, suggesting that the follower did not have any significant bias. The results of the simulations are plotted in red along with the corresponding experimental results in Fig. 4 (d), (e), (g) and (h).

There are two findings from the above experiments that are of critical importance for large networks of interconnected p-bits:

1. A weak electrical interconnection, which is more than 10 times smaller current than that required for deterministic switching, is sufficient to induce correlations between two p-bits. Weak interconnection strength is crucial for low power consumption in a large network. Moreover, as correlations are present even with weak interconnection strengths, it allows for electrical annealing[7], where the interconnection strength can be gradually turned up to further enhance the desired correlations and suppress the undesired ones.
2. A large difference in the natural time scales of the two devices does not hamper the operation of such circuits. Another important factor for a large p-bit network is its robustness against device to device variations. Specifically, the natural fluctuation time scales of the p-bits depend exponentially on the energy barrier of the individual nanomagnet, which can have a distribution due to process variability. Therefore, for a network of p-bits with different energy barriers to work as desired, correlations need to be established even with different individual fluctuation time scales. This important requirement has been verified in our experiments, where correlations were successfully established between the two p-bits despite their natural time scales being very different (24 ms and 648 ms for driver and follower respectively).

## Conclusion

In summary, we have demonstrated for the first time, a stochastic nanomagnet with perpendicular anisotropy, tunable by an in-plane spin current. We hypothesize the possibility of a tilted anisotropy being responsible for the observed tunability, which is supported by both experiments and sLLG simulations. We further demonstrate a coupled network of two such stochastic devices, namely p-bits, and show that correlations between their stochastic outputs can be manipulated through weak electrical interconnections, despite having difference in their natural fluctuation time scale.

## Acknowledgement

The authors would like to thank Prof. Joerg Appenzeller and Prof. Pramey Upadhyaya for many fruitful discussions and their feedback on this work. P.D. would like to thank Vaibhav Ostwal, Terry Hung and Tingting Shen for their helpful feedback during the conduction of this work.

This work was supported by the Center for Probabilistic Spin Logic for Low-Energy Boolean and Non-Boolean Computing (CAPSL), one of the Nanoelectronic Computing Research (nCORE) Centers as task 2759.003 and 2759.004, a Semiconductor Research Corporation (SRC) program sponsored by the NSF through CCF 1739635.

# Supplementary Information

This supplementary text contains two sections to characterize our p-bit as a random number generator:

(i) Dependence of the speed of random bit generation on the thermal stability factor of the nanomagnet
(ii) Evaluation of the quality of the generated random bits according to the statistical test suite provided by the National Institute of Standards and Technology (NIST)

## Characterizing the device as a high quality random number generator

The telegraphic output obtained from our devices are digitized to obtain a random bit stream. This bit stream can be used as a source of high quality random numbers, as required by many applications. We study speed of random bit generation by our devices and provide strategies to improve it. Then we evaluate the quality of the generated random bit stream using standard test protocols according to the statistical test suite (STS)[SR1] provided by the National Institute for Standards and Technology (NIST).

*Speed of random bit generation*

The obtained random bit stream is used to generate the histogram plot of the dwell time in the two metastable states as shown in Fig. 2 (c). An exponential envelope fits well with the experimentally obtained dwell time histogram, suggesting that the random fluctuations of nanomagnets can be well described by a Poisson process. Average dwell time in the "UP" and "DN" states, i.e. $\tau_{UP}$ and $\tau_{DN}$ are obtained from the exponential fitting. For a completely symmetrical energy landscape, these two dwell times should be identical. However, we obtain a slightly skewed distribution due to the remanent magnetic field in the measurement setup. The speed of the random bit generation can be obtained from the harmonic mean of $\tau_{UP}$ and $\tau_{DN}$ ($\tau^{-1} = \tau_{UP}^{-1} + \tau_{DN}^{-1}$). This time scale is determined by the energy barrier of the nanomagnet through:

$$\tau = \tau_0 \times \exp(E_B/k_B T) \quad \ldots (2)$$

Hence, the speed of the random number generation can be increased by reducing the thermal stability factor, $E_B/k_B T$. From the obtained $\tau$, we determine $E_B/k_B T \approx 18$ for the stochastic device presented here. Fig. S1 shows the experimentally measured effect of changing this stability factor by changing temperature. The fluctuation time scale, $\tau$, changes by more than 3 orders of magnitude. This stability factor can be reduced in practice by scaling the volume of the nanomagnet further to increase the speed of random number generation.

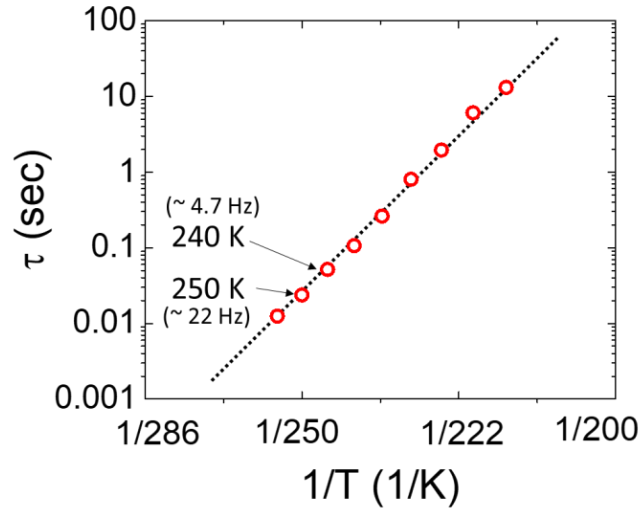

Fig. S1: Speed of random number generation can be changed by changing the thermal stability factor $E_B/k_BT$. To show this, we measure the telegraphic output at different temperatures. This plot shows the Arrhenius plot of the fluctuation time scale, $\tau$, of the magnet as a function of temperature. The time scale can be changed by more than 3 orders of magnitude by changing the temperature by 50 K.

*Quality of the generated random bit stream*

We evaluate the quality of the random bits generated from our device using the STS provided by NIST. Various aspects of the generated random sequence are tested for qualifying as cryptographic quality random numbers. These tests compute the statistics of the input sequence such as mean value, standard deviation, entropy, repeated structures, linear dependencies, autocorrelation, etc. and compare them against theoretical expectations from a perfectly random sequence. Details of the tests can be found in the manual of the NIST STS package. Prior to performing the tests, we do the following post-processing of the collected data in order to reduce any present bias. After digitizing the data, we take the XOR of the bit stream with a copy of itself that is shifted by one time step. This is shown in Fig. S2. The resulting p-values of the NIST test are given in table 1. A p-value >0.01 means the test has been passed. The generated bit stream by our device passed all 9 tests that were performed, showing cryptographic quality randomness. Some of the tests required > $10^6$ bits and were not performed due to the limited size of our data set (10,000 bits).

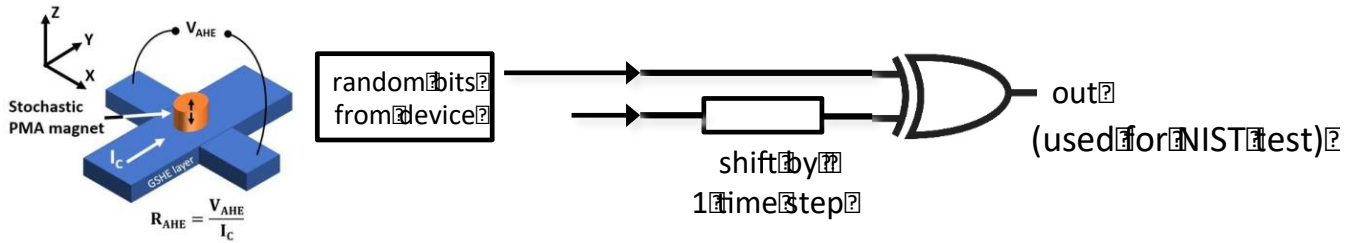

Fig. S2: Post processing of the output bit stream prior to performing NIST randomness test. The obtained random bits are XORed with a time shifted version of itself to remove any bias.

|    | Test Name                 | Description                                         | p-value         |
|----|---------------------------|-----------------------------------------------------|-----------------|
| 1  | Frequency                 | Total number of 0's and 1's mismatch                | 0.3764          |
| 2  | Block frequency           | Number of 0's and 1's mismatch within a subset      | 0.6526          |
| 3  | Runs                      | Too many sequences of consecutive bits of one type  | 0.9693          |
| 4  | Longest run               | Too many consecutive bits of one type               | 0.6600          |
| 5  | Binary matrix rank        | Linear dependence among fixed length substrings     | 0.4613          |
| 6  | Discrete Fourier transform| Periodicity in bit stream                           | 0.9884          |
| 7  | Serial                    | Multiple low entropy sequences in a row             | 0.0344, 0.5958  |
| 8  | Approximate Entropy       | Bit sequence too unlikely                           | 0.0601          |
| 9  | Cumulative Sums (forward) | Running sum deviates too far from half the length   | 0.2143          |
| 10 | Cumulative Sums (backward)| Same as previous, but in reverse direction          | 0.6776          |

Table 1: Results of the NIST STS tests on the pre-processed random bit stream. P-value>0.01 means that the test has been passed according to the 1% confidence level.